\documentclass[twocolumn,showpacs,prl]{revtex4}
\usepackage{graphicx}
\begin{document}

\title{Structural studies of phosphorus induced dimers on Si(001)}
\author{Prasenjit Sen}
\affiliation{Harish-Chandra Research Institute \\ Chhatnag Road,
Jhunsi, Allahabad 211019, INDIA.}
\author{Bikash C.\ Gupta and Inder P.\ Batra}
\affiliation{University of Illinois at Chicago \\ 845 W. Taylor
Street, Chicago Illinois 60607, USA.}

\begin{abstract}
Renewed focus on the P-Si system due to its potential application 
in quantum computing
and self-directed growth of molecular wires, has led us to study 
structural changes 
induced by P upon placement on Si(001)-$p(2\times 1)$.
Using first-principles density functional theory (DFT)
based pseudopotential method,
we have performed calculations for P-Si(001) system, starting from
an isolated P atom on the surface, and systematically increasing the 
coverage up to a full monolayer.
An isolated P atom can favorably be placed on an {\bf M} site
between two atoms of adjacent Si dimers belonging to the same Si dimer
row. But being 
incorporated in the surface is even more energetically beneficial due
to the participation of the {\bf M} site as a receptor for the ejected Si. 
Our calculations show that up to $\frac{1}{8}$ monolayer  
coverage, hetero-dimer structure resulting from replacement of
surface Si atoms with P is energetically favorable.
Recently observed zig-zag features in STM are found to be consistent
with this replacement process. As coverage increases, 
the hetero-dimers give way to P-P ortho-dimers on the Si dimer rows.  
This behavior is similar to that of Si-Si d-dimers
but are to be contrasted with the Al-Al dimers, which 
are found between adjacent Si dimers rows and in a para-dimer arrangement. 
Unlike Al-Si system
P-Si does not show any para to ortho transition.  For both systems, 
the surface reconstruction is lifted at about 
one monolayer coverage.  These calculations help us in 
understanding the experimental data obtained using 
scanning tunneling microscope.
 
\end{abstract}
\maketitle

\section{Introduction}

Phosphorous-doped Si is the back-bone of micro-electronic technology.
Recently, P-Si system has generated renewed interest due to its
potential application in quantum computers~\cite{obrien,schofield,kane} and
growth of molecular wires on Si surfaces~\cite{wangwire}. It is also of
fundamental interest to compare the behavior of P-P dimers with Si-Si
and Al-Al dimers on the Si surfaces, a lot having been understood about 
the last system~\cite{brockss,brocksal,bikash}.

Phosphine gas (PH$_3$) is used as the source of P in most applications.
Yu {\it et al.}~\cite{yu3} concluded from their experiments that 
around 675$^0$ C, all
hydrogen atoms from PH$_3$ are desorbed and the surface is a monolayer
(ML) P covered
Si(001) with the formation of P-P dimers. Wang {\it et al.}~\cite{wang} did a
scanning tunneling microscopy (STM) and Auger electron spectroscopy
(AES) study of phosphorous-terminated Si(001) surface 
close to a monolayer P coverage. They find
mostly P-P dimers, along with some
Si-P dimers on the surface. They also find defects in the P-P dimer
rows as well as anti-phase boundaries. At slightly below a ML P
coverage, Wang {\it et al.} observe Si-Si, Si-P and P-P dimers. At still lower
coverages, there are `significant' numbers of Si-Si and Si-P dimers,
while there are some P-P dimers. Kipp {\it et al.}~\cite{kipp} using STM, X-ray
photoemission spectroscopy (XPS) and total energy calculations
conclude that after low temperature PH$_3$ adsorption there are dimers
on the surface. It is not conclusive whether
these are P-P, Si-P or Si-Si dimers, though they expect P-P dimers to be
dominant at low temperatures. Both these groups (Wang
{\it et al.} and Kipp {\it et al.}) observe similar `bright' features
above Si-Si surface dimers in their STM images at low P coverages.
Wang {\it et al.} claim these to be
indicative of Si-Si dimers, while Kipp {\it et al.} claim these to be
P-P dimers. Curson {\it et al.}~\cite{curson}, from their STM studies, 
conclude that at low P coverages, the surface, in fact, contains Si-Si
and Si-P dimers, thus supporting Wang {\it et al.}'s conclusions. 
Kipp {\it et al.} also found that 
after PH$_3$ adsorption
at 625$^0$ C, there are only symmetric P-P dimers on the surface. At
the maximum P coverage, they find defects like missing dimer rows. From
thermal desorption spectra of P from the Si(001) surface, Hirose and
Sakamoto~\cite{hirose} claim that at low coverages ($< 0.2$ ML), there
are mostly Si-P dimers on the surface. Above $0.2$ ML coverage, P-P,
Si-P dimers and defects coexist on the surface. Lin {\it et
al.}~\cite{lin} in their
core-level photoemission and STM studies find that
at $\sim 700$ K the surface %is a hydrogenated Si(001) surface 
is interspersed with chains of P-P dimers. At $\sim 800$ K
the hydrogen atoms desorb completely and one observes partial
replacement of Si atoms by P. 

Thus apart from some differences in the  details, most experiments 
agree that after the hydrogen from PH$_3$ has been desorbed, 
the surface consists mostly of P-P dimers at higher coverages.
At low coverages there is agreement that P replaces Si atoms and gets
incorporated into the surface, but
one would like to have a more detailed understanding of the structure.

There have been a few theoretical studies addressing the
question whether PH$_3$ adsorbs dissociatively or molecularly on
Si(001)~\cite{miotto,cao,wilson}. 
To our knowledge, there are no systematic theoretical studies of P-covered
Si(001) surface as a function of coverage.
Since it is the P atoms which are important
in applications, one would like to understand their interaction with and 
structure on the Si(001) surface.

Therefore, we study stable atomic
structure of P-covered Si(001) surface starting from a low coverage
up to a ML. Apart from finding the most favorable binding site for a
P ad-atom on this surface, we study P-P dimers on 
Si(001) in detail because they turn out to be a more favorable
arrangement compared to isolated P atoms. We also compare the
structure and energetics of P-P ad-dimers with Al-Al ad-dimers about
which much is known.
We arrive at the important conclusion that unlike Al-Si system, P-Si
system always prefer an ortho-dimer structure and does not show a
para to ortho transition with increasing coverage. 
Since P-Si hetero-dimers have been observed in experiments, we have also
studied the energetics of replacement of surface Si atoms with P. In
fact, at low P coverages, this replacement of
surface Si atoms is a more favorable arrangement than adsorption of
the P ad-atoms or dimers above the surface. Interestingly, the bright lines and
zig-zag features observed in ref.~\cite{curson} are related to this
replacement process.
In what follows, we discuss the methods used, and the
results of our calculations in detail.

\section{Method}
\label{sec:method}

Calculations were performed using pseudopotential method within 
DFT. We use VASP~\cite{vasp0,vasp1}
for our calculations. The wave functions are expressed in a 
plane wave basis with an energy cutoff of $250$ eV. The Brillouin zone
integrations are performed using the Monkhorst-Pack scheme~\cite{monkho}.
Ionic potentials
are represented by ultra-soft Vanderbilt type pseudopotentials~\cite{vander}.
We use
the generalized gradient approximation (GGA)~\cite{perdew} for the
exchange-correlation energy. The preconditioned conjugate gradient
method (as implemented in VASP) is used for wave function optimization
and conjugate gradient is used for ionic relaxation. 
We use a $(2\times2\times1)$ \textbf{k}-point mesh for our 
$(4\times 4)$ surface supercell, while for the $(2\times 2)$ 
surface supercell, we use a $(4 \times 4\times 1)$ \textbf{k}-point mesh.
Convergence with respect to energy cutoff and number of \textbf{k}
points has been previously tested for similar
systems~\cite{site,tony}.
When making comparison between 
binding energies of structures with same supercell size, it is
expected that the errors due to cutoff and \textbf{k}-point mesh will
cancel. 

The $\mathrm{Si(001)-\mathnormal{p}(2\times1)}$ 
surface is represented
by a repeated slab geometry. Each slab contains 5 Si atomic layers
with hydrogen atoms passivating the Si atoms at the bottom layer of
the slab. Consecutive slabs are isolated from each other by a vacuum
space of 9 \AA. The Si atoms in the top four
atomic layers are allowed to relax, while the bottom layer Si atoms
and passivating H atoms are kept fixed in order 
to simulate bulk-like termination.
We reproduced the energetics and geometry of the $p(2\times1)$ reconstructions
of a clean Si(001) surface using the above parameters~\cite{kelly0}.

Our calculations provide cohesive energy of a supercell composed
of given set of atoms,
\begin{equation}
E_{C}[SC]=E_{T}[SC]-E_{A}[atoms]\label{eq:EC}
\end{equation}
where $E_{C}[SC]$ is the cohesive energy of the supercell, $E_{T}[SC]$
is the total energy
of the supercell, and $E_{A}[atoms]$ is the total energy of all the isolated
atoms that constitute the supercell. Thus $E_{C}[SC]$ is the energy
gained by assembling the given supercell structure from the
isolated atoms. We define the binding energy (BE) of $n$ P atoms, $E_{B}$ as, 
\begin{equation}
E_{B}=E_{C}[\mathrm{Si}]-E_{C}[\mathrm{Si+\mathnormal{n}P}]
\label{eq:EB1}
\end{equation}
where $E_{C}\mathrm{[Si]}$ is the cohesive energy of the 
Si slab, and $E_{C}\mathrm{[Si+\mathnormal{n}P]}$ is the cohesive
energy with $n$ P atoms adsorbed/incorporated into the slab.
The cohesive energies of the Si slab with and without P are calculated
in the same supercell with fully relaxed atomic configurations. 
Written in terms of the total energies, it is easy to see from
eqn.~\ref{eq:EC}, that the BE can be expressed as,
\begin{equation}
E_B = E_T[\mathrm{Si}] - E_T\mathrm{[Si+\mathnormal{n}P]} + n
E_A[\mathrm{P}]
\label{eq:EB2}
\end{equation}

It should be noted that in order to compare stabilities of two structures,
one should compare their formation energies (FE), which, in case of an
`interstitial' impurity (in this case, added P atoms may replace Si atoms in
the slab, but they all remain within the system) can be written
as~\cite{dalpian}
\begin{equation}
E_{form} = E_T[\mathrm{Si+\mathnormal{n}P}] - E_T[\mathrm{Si}] - n\mu_P 
\label{eq:FEi}
\end{equation}
\noindent where $E_T[Si+nP]$ is the total energy of the Si slab with
the $n$ P atoms, $E_T[\mathrm{Si}]$ is the total energy of the Si slab, 
and $\mu_P$ is the chemical potential of phosphorous
in its reference state. In the case when different structures being
compared have equal number of Si and P atoms, 
one can see from eqns~\ref{eq:EB2} and ~\ref{eq:FEi} 
that comparing their FE's
is equivalent to comparing their BE's, since difference in FE's is just the
negative of the difference in BE's. Also, if the reference state is taken
to be a gas of isolated atoms, which is probably appropriate in MBE
conditions, then the FE of a structure is just the negative of its BE.
However, in case of a `substitutional' impurity, when a Si atom replaced by
a P atom leaves the system and goes to a reservoir 
(again assumed to be a gas of
isolated Si atoms), the formation energy is given by~\cite{dalpian}
\begin{eqnarray*}
E_{form} & = & E_T[\mathrm{(N-1)\; atom\; Si\; slab + P}] \\
         & - & E_T[\mathrm{N\; atom\; Si\; slab}] \\
         & + & E_A[\mathrm{Si}] - E_A[\mathrm{P}]
\label{eq:FEs}
\end{eqnarray*}
Again, from eqn~\ref{eq:EC}, it is easy to see that this is equal to the
difference between the cohesive energies of an $N$-atom 
Si slab in which one Si is replaced by a P atom, and a the clean $N$-atom
Si slab.  

In this work, we mostly study
stabilities of various structures having the same number of Si and P atoms,
as suggested by experiments. Hence comparing their BE's serves the purpose, 
higher BE implying a more stable structure. In case of an isolated P
being incorporated in the Si slab, we also compare stabilities of these two
structures: 1. the ejected Si remains in the slab; and 2. it goes to a
reservoir. For this, we do have to compare the FE's, as discussed in the
next section.

\section{Results and Discussions}

In the subsequent subsections, we present results of our calculations 
in detail. We calculate the energetics of P adsorption on the surface, and
replacement of surface Si atoms by P when we have an isolated P
ad-atom, and at
$\frac{1}{8}$, $\frac{1}{4}$, $\frac{1}{2}$ and a full ML coverages.
\label{sec:sparse}

\subsection{Isolated P ad-atom}

\begin{figure}
\scalebox{0.50}{\includegraphics{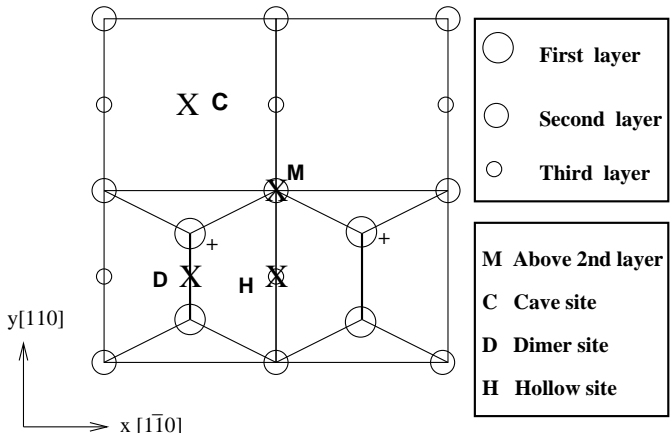} }
\hfill
\caption{Symmetry sites on the $p(2\times 1)$ asymmetric Si(001)
surface at which binding properties of an isolated P ad-atom are
studied. The Si atoms marked `$+$' are at a greater height compared to
their partners in the same dimer. This system size is for illustration only. 
Calculations have been done on different system sizes as discussed in
the text.} 
\label{fig:sites}
\end{figure}

As stated earlier, for P atoms on the Si(001) surface, we have considered two
possibilities: they get adsorbed on the surface or they replace 
Si atoms and get incorporated into the surface. 
In the first case, we have calculated the BE of a single 
P ad-atom adsorbed at four
symmetry sites on the $p(2\times 1)$ asymmetric Si(001)
surface. The symmetry sites are: i) dimer site ({\bf D})
on top of the Si surface dimer, ii) the site
vertically above the second
layer Si atom between two atoms of adjacent Si dimers
belonging to the same dimer row ({\bf M}), iii) cave site ({\bf C}) 
between two Si surface dimers perpendicular to the dimer rows
and iv) quasi-hexagonal site ({\bf H}) 
half-way between two Si surface dimers along
a dimer row. These sites are marked in fig~\ref{fig:sites}.
We used a $(4\times 4)$ surface supercell. The large size of the
supercell ensures that the interaction between a P atom in our supercell
and its periodic images are small and the BE represents that of an
isolated P atom. While studying BE of a P ad-atom at these sites, we
relax the P atom, and the top four Si layers. 
The BE values of the P atom at these sites are
given in table~\ref{table:BE164ML}.
In table~\ref{table:BE164ML} we also give BE of one P atom in a
$(2\times 2)$ surface cell corresponding to $\frac{1}{4}$ ML, which is
discussed in detail later.

\begin{figure}[h]
\scalebox{0.5}{\includegraphics{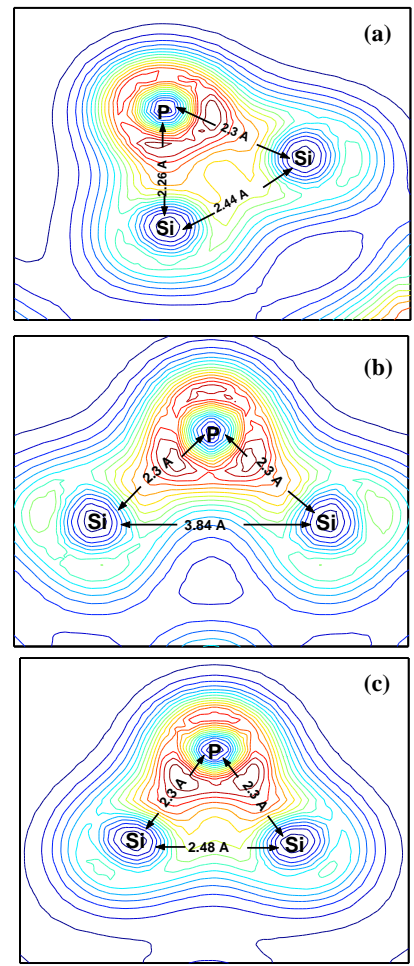}}
\caption{(Color online) 
Charge density contour plots in the plane of the P ad-atom
and the Si atoms it binds with. (a) P ad-atom at the {\bf M} site
bonded to a second layer Si atom and a surface Si atom;
(b) P ad-atom at the {\bf M} site bonded to two surface Si atoms of
adjacent dimers in the same dimer row; 
(c) P ad-atom at the {\bf D} site bonded to two Si atoms of the same surface
dimer.}
\label{fig:SD_charge}
\end{figure}

The {\bf M} site turns out to be the most favorable site for
an isolated P ad-atom. This is also the most favorable site for an
isolated Si ad-atom on Si(001) first discovered by Brocks {\it et
al.}~\cite{brocksSi}. Here P ad-atom binds to two Si atoms belonging
to two different dimers in the same surface dimer row.
These bonds are of equal length (2.3 \AA) with Si-P-Si bond
angle $\sim 112^0$, suggesting that P likes to be close to a
tetrahedrally bonded configuration. Another significant observation
is that the P ad-atom at the {\bf M}-site is only 2.26 \AA~ away from
the second layer Si atom. This is a bond similar in character to the Si-P
bonds at the surface as seen in the charge density plot in
fig.~\ref{fig:SD_charge}(a). Thus the
second layer Si atom bonded to P becomes five-fold coordinated
probably accompanied by
weakening of its bonds with other Si atoms. This
weakening of bonds costs energy, but P being 3-fold coordinated 
at the {\bf M} site is beneficial energetically compared to the {\bf D}
site, where the P ad-atom is only two-fold coordinated.
It is also found that 
when the P atom is at the {\bf D} site, the Si-P-Si bonds make an
angle of 64$^0$. 
This angle is much smaller than an ideal tetrahedral angle of 109$^0$.
So there is a bond rotation on the P atom at the {\bf D} site which
costs energy. An interplay of these factors makes the {\bf M} site
more favorable by 0.2 eV compared to the {\bf D} site.
We show the charge density contour plots in the plane of the P ad-atom
and the two surface Si atoms it binds with at the {\bf M} and {\bf D}
sites in fig~\ref{fig:SD_charge}(b) and (c).
The nature of P-Si bonding at the two sites is similar, so it is the
bond rotation at the {\bf D} site, and rearrangement of bonds the second 
layer Si atom forms, that make the {\bf M} site more favorable.

Energetically, {\bf M} and {\bf D} sites are followed by the {\bf H} and {\bf C}
sites.
At the {\bf H} and {\bf C} sites, the P ad-atom can bind to four and 
two surface Si atoms respectively. 
However, being an $sp$ element, and having a small
atomic radius, it cannot take full advantage of
all the neighboring surface Si atoms. Hence {\bf H} site turns
out to have a lower BE than the {\bf M} and {\bf D} sites. 
The {\bf C} site has the lowest BE. 
It has been seen before that the size of an ad-atom can have dramatic
effects on its binding properties on a substrate~\cite{brockss}. Thus
while the {\bf M} site was found to be the most favorable
site as stated before, for an Al ad-atom, {\bf H} site turned out to the most
favorable~\cite{brocksSi,brocksal}.

\begin{table}[h]
\caption{BE (eV/atom) for an isolated P ad-atom at different symmetry 
sites on a Si(001)-$p(2\times 1)$ surface at two different coverages.}
\vspace*{0.1in}
\begin{tabular}{ccc}
\hline
\hline
site \hspace*{0.5in} & isolated P \hspace*{0.5in} & $\frac{1}{4}$ ML \\
\hline
{\bf M} \hspace*{0.5in} & 5.75 \hspace*{0.5in} & 5.73    \\
{\bf D} \hspace*{0.5in} & 5.55 \hspace*{0.5in} & 5.55  \\
{\bf H} \hspace*{0.5in} & 4.76 \hspace*{0.5in} & 4.74 \\
{\bf C} \hspace*{0.5in} & 3.73 \hspace*{0.5in} & 4.45 \\
\hline
\end{tabular}
\label{table:BE164ML}
\end{table}

\begin{figure}[h]
\scalebox{0.5}{\includegraphics{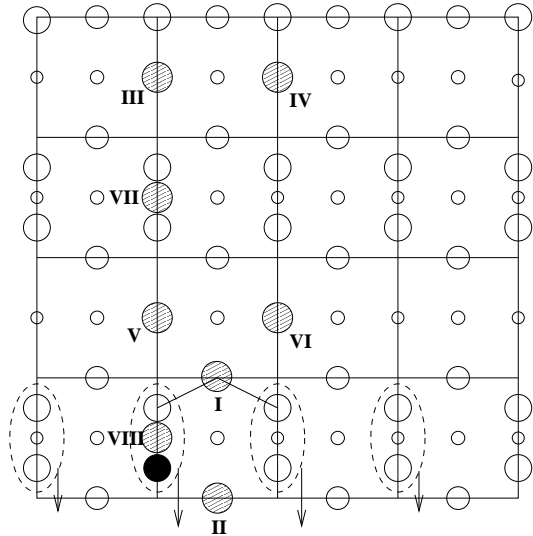}}
\caption{Starting atomic geometries for studying incorporation of one P
atom into a $(4 \times 4)$ surface cell. The incorporated P atom is
shown in dark color. Shaded circles show possible positions of the
ejected Si atom, while open circles are the Si atoms in the slab. 
The movements of the Si-P dimer and Si-Si dimers
neighboring it are indicated by the arrows.}
\label{fig:1Pinc}
\end{figure}

When the P atom gets incorporated into the
surface we replace one of the Si atoms in a surface dimer on a
$(4\times 4)$ cell by P. As for the ejected Si atom, it can either go to a
reservoir, or bind with the Si surface at some other site. Experiments
support the latter scenario~\cite{curson,wilson}. However, for the sake of
completeness, we have compared the energetics of these two scenarios, as
discussed later. In cases where it remains in the system,
we place the ejected Si atom at various sites on the
surface. The possible initial geometries are shown in
fig~\ref{fig:1Pinc}. Starting from these geometries, the ejected Si atom, 
top four atomic layers
(including the incorporated P) are relaxed in all directions.
It turns out that the energetically most favorable
position for the ejected Si atom is geometry 
{\bf I} marked in fig~\ref{fig:1Pinc}. This agrees with the results of
Wilson {\it et al.}~\cite{wilson}. 
In the final relaxed geometry, the Si-P hetero-dimer moves by 
$\approx 0.6$ \AA~ 
along $\langle \bar{1}\bar{1}0 \rangle$ compared to the
Si-Si dimers on the clean surface. 
The adjacent Si-Si dimer, on the side of the ejected Si, moves by
$\approx 0.35$ \AA~ along $\langle \bar{1}\bar{1}0 \rangle$. 
The two Si-Si dimers neighboring these two dimers also move by
$\approx 0.1$ \AA~ along $\langle \bar{1}\bar{1}0 \rangle$.
As noted, this movement of the dimers relative to the original surface
dimers propagates along the dimer row up to at
least 2 dimers away from the Si-P dimer (the maximum distance observable
in a $4\times 4$ cell). The BE of the P atom in geometry geometry {\bf I} 
is 6.1 eV.  The BE's of the P atom in all the geometries studied are given in 
table~\ref{table:BE1Pinc}. 
Most importantly, comparing two situations with an isolated P atom
on the surface: (i) when the P atom is adsorbed at the {\bf M} site, and
(ii) when it forms a P-Si heterodimer with the ejected Si in geometry
{\bf I}, the latter case is found to be more favorable by 0.4 eV.
It is interesting to note that geometry {\bf I} is reminiscent of Si
ad-atom having {\bf M} as the most favorable site~\cite{brocksSi}. The
participation of {\bf M} site makes Si replacement a favorable
scenario. This is consistent with the observation that at
low coverages, P atoms get incorporated into the Si surface.

We now ask whether it is more favorable for the ejected Si to remain in
the system, or to go to a reservoir. For this, we compare the FE's of the
following structures: 1. geometry {\bf I} discussed above; 
2. a P atom replacing a Si atom in
a surface dimer, and the ejected Si going to a reservoir that is assumed to
be a gas of Si atoms. As argued in the METHOD section, the FE in the
first case is just the negative of its BE, {\i.e.}, $-6.1$ eV. The FE in the
second case, calculated as discussed in eqn.~\ref{eq:FEs},
turns out to be $-1$ eV. Since higher FE means a less stable structure, 
clearly, it is more favorable for
the ejected Si to remain within the system, a conclusion that matches with
experimental observations. In all subsequent discussions, we assume that
all the ejected Si atoms remain in the system.

\begin{table}[h]
\caption{BE in eV per P atom incorporated into the Si(001) surface 
with the ejected Si atom at different sites as discussed in the text.}
\vspace*{0.1in}
\begin{tabular}{cr}
\hline
\hline
geometry \hspace*{0.5in} & BE         \\ \hline
\hline
{\bf I} \hspace*{0.5in}      & 6.1 \\
{\bf II} \hspace*{0.5in}      &  5.2 \\
{\bf III} \hspace*{0.5in}      & 5.4 \\
{\bf IV} \hspace*{0.5in}      & 5.4 \\
{\bf V} \hspace*{0.5in}      & 4.9 \\
{\bf VI} \hspace*{0.5in}      & 5.2 \\
{\bf VII} \hspace*{0.5in}      & 5.4 \\
{\bf VIII} \hspace*{0.5in}   & 5.1  \\
\hline
\label{table:BE1Pinc}
\end{tabular}
\end{table}

\subsection{$\frac{1}{8}$ ML P Coverage}

At a $\frac{1}{8}$ ML coverage, again we consider two
possibilities--P atoms replacing surface Si atoms,
or they being adsorbed on the surface.

\begin{figure}[h]
\scalebox{0.45}{\includegraphics{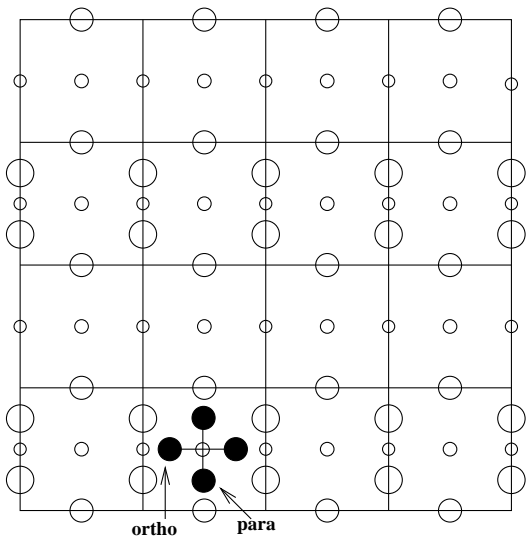}}
\caption{The para- and ortho- orientations of a P-P dimer on the Si(001)
surface at $\frac{1}{8}$ ML studied in this work.}
\label{fig:44_1dimposs}
\end{figure}

When the P atoms are adsorbed on the surface, we put 2 P ad-atoms on a
$(4\times 4)$ surface supercell. Phosphorous
being a pentavalent atom, even after
binding to one or two surface Si atoms, we expect two P ad-atoms to
dimerize if possible. This is borne out by our calculations at
$\frac{1}{4}$ ML, as we discuss later. Hence, at
$\frac{1}{8}$ ML, we consider possible positions of a P-P dimer. 
The {\em para-} and {\em ortho-dimer} arrangements are shown in
fig.~\ref{fig:44_1dimposs}.
We find that the ortho-dimer is
a more favorable configuration of the P-P dimer, with a BE of 6.3 eV
per P atom. The binding of the para-dimer turns out to be 
weaker with a BE of 5.9 eV per P atom. 

It is interesting to compare these energetics with those of Al-Al
dimers on the Si(001) surface. The biggest difference between P-P
and Al-Al dimers on Si(001) is that while Al-Al dimers prefer
to reside between surface dimer rows~\cite{bikash,brocksal}, 
P-P dimers find it favorable
to adsorb on top of dimer rows. We did study a P-P para-dimer in between
two surface dimer rows, and the binding is found to be substantially weaker
with a BE of 5.4 eV per P. Also, in the case of Al-Al dimers, at low
coverages, the para-dimer configuration was found to be more
favorable, while for P, ortho-dimer is more favorable.
On the other hand, for a Si-Si ad-dimer on Si(001),
Brocks {\it et al.} found that a para-dimer on
a surface dimer row is only slightly more favorable by 0.1 eV compared to an 
ortho-dimer~\cite{brockss} (as this energy difference is the
limit of the accuracy of their calculations). 
As we have found, a P ad-atom at the {\bf
M} site forms a bond with the second layer Si atom directly below it. This bond
length is, in fact, slightly shorter than the bonds the P ad-atom
makes with the surface Si atoms(see fig~\ref{fig:SD_charge}). 
This causes weakening of bonds
between the second layer Si and other Si atoms, as we have already
argued. One can view a para-dimer being formed by dimerization of
two P ad-atoms on two {\bf M} sites between two Si-Si dimers in the
same surface dimer row. This stretches the P-second layer Si bonds, 
while the bonds between the second
layer Si and other Si atoms are still weak. The overall effect is a
net energy cost. In the ortho-dimer orientation, the P ad-atoms do not
affect any second layer Si atoms, while they still dimerize and bind
with four surface Si atoms. This situation turns out to be more
favorable. In case of a Si ad-atom at the {\bf M} site, the distance
between the Si ad-atom and the second layer Si is found to be
greater than bulk Si-Si distance, and also greater than 
the distance between the Si ad-atom and surface Si atoms~\cite{brocksSi}. 
Hence a Si ad-atom at the {\bf M} site
has negligible effect on the second layer Si atom. Thus even in a
para-dimer orientation, there is not much energy cost in stretching the
ad-atom-second layer bonds, and, in fact, para-dimer becomes more
favorable for Si ad-dimers.

\begin{figure}
\scalebox{0.35}{ \includegraphics{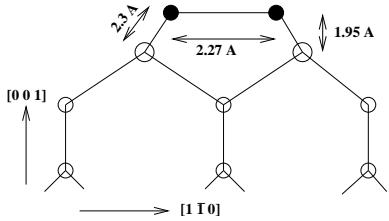} }
\caption{Side-view of the local geometry around a P-P ortho-dimer adsorbed on
Si(001). The P-P dimer is found to be symmetric.}
\label{fig:1dim}
\end{figure}

A side-view of the P-P ortho-dimer on the Si(001) surface is
shown in fig~\ref{fig:1dim}. The P ad-atoms symmetrize
the two adjacent Si-Si surface dimers, but do not break the dimers.
Phosphorous atoms having
smaller atomic radii than Si atoms, the P-P bond length is smaller ($\sim
2.27$ \AA). 
Hence the two adjacent Si-Si dimers are
drawn in closer to the P-P dimer along $\langle 1\bar{1}0 \rangle$ 
direction compared
to their position on the $p(2\times 1)$-asymmetric
reconstructed surface.
The Si-P distance in this case turns out to be \mbox{$\sim 2.3$ \AA}.

\begin{figure}[h]
\scalebox{0.4}{\includegraphics{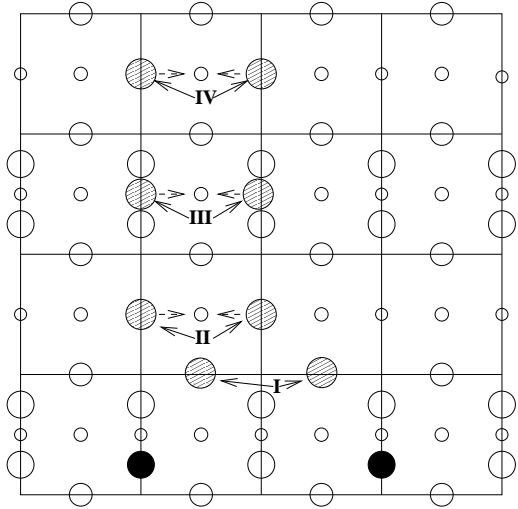}}
\caption{Starting atomic arrangements for incorporation of two P atoms
into $(4\times 4)$ surface cell. The P atoms are denoted by dark circles. 
Shaded
circles show possible positions of the pair of Si atoms displaced by the
P atoms. Small arrows indicate that the two ejected Si atoms form a dimer.} 
\label{fig:2Pinc}
\end{figure}

When two P atoms are incorporated,
the ejected Si atoms are placed at various sites on the surface and
the geometries studied in this paper are shown in fig~\ref{fig:2Pinc}. 
Starting from these
geometries, the ejected Si atoms, and top four atomic layers are relaxed.
After relaxation, geometry {\bf III}
turns out to be the most favorable with a BE of 6.4 eV per P atom.
Note that geometry {\bf III} is an ortho- orientation of the Si-Si
dimer on top of a surface dimer row. But it is known that the energy
difference between the para- and ortho- configurations is small ($\sim
0.1$ eV)~\cite{brockss}, and hence either structure may be seen in
experiments. In the final converged geometry, the two ejected Si atoms form an
asymmetric dimer with a Si-Si distance of 2.39 \AA.
In this geometry the dangling bonds of the four surface Si atoms,
above which the ejected Si atoms dimerize, are saturated. The ejected Si atoms
are also 3-fold coordinated. This large reduction of dangling bonds causes
this geometry to be the most favorable. In geometry {\bf I}, the next most
favorable arrangement, the ejected Si
atoms 
do not dimerize. In this geometry,
four dangling bonds of the surface Si atoms are
saturated, but the ejected Si atoms are only 2-fold coordinated.
In geometries {\bf II} and {\bf IV} also, after relaxation, the ejected Si
atoms form asymmetric dimers with Si-Si dimer distances of 2.38 \AA~ and
2.41 \AA~ respectively. However, in these two geometries, the surface Si
atoms are quite far from the ejected Si's. Thus there is not much energy
gain from bonding with surface Si atoms. This causes {\bf II} and {\bf IV}
to be lower in BE.
The BE's for all the geometries are given in table~\ref{table:BE2Pinc}.

\begin{table}[h]
\caption{BE per P atom for P incorporated into the Si(001) 
surface at $\frac{1}{8}$ ML. The geometries refer to various positions
of the ejected Si atoms as discussed in the text.}
\vspace*{0.1in}
\begin{tabular}{cr}
\hline
\hline
geometry \hspace*{0.5in} & BE         \\ \hline
\hline
{\bf I} \hspace*{0.5in}      &   6.0  \\
{\bf II} \hspace*{0.5in}      &  5.6 \\
{\bf III} \hspace*{0.5in}      & 6.4 \\
{\bf IV} \hspace*{0.5in}      & 5.7 \\
\hline
\label{table:BE2Pinc}
\end{tabular}
\end{table}

Again, what is interesting to note is that the geometry with P atoms
incorporated into the surface has a lower energy. 
%(Compare tables~\ref{table:BE1dim} and ~\ref{table:BE2Pinc}). 
Therefore, at low P
coverages, it is more favorable for P atoms to replace surface Si
atoms. The ejected
Si atoms prefer to go in positions where the next layer of Si atoms
would be above the starting surface, and form asymmetric dimers. This
suggests that the bright lines seen in the STM images, along
with the Si-P heterodimers are, in fact, ejected 
Si-Si dimers, and not P-P dimers. 
This is consistent with Curson {\it et al.}'s interpretation of
the lines perpendicular to the surface dimer rows in their STM images
as Si-Si dimer chains~\cite{curson}. 
This also supports Wang {\it et al.}'s interpretation of their STM 
images~\cite{wang}.

\begin{figure}[h]
\scalebox{0.55} { \includegraphics{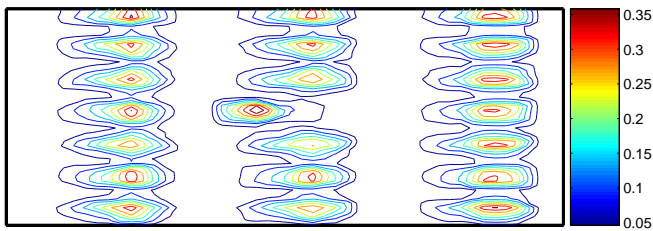} }
\caption{(Color online)
Charge density contour plots in a plane $\sim 1$ \AA~ above 
the Si-P dimer. The displacement of the Si-P dimer gives it a zig-zag
appearance in STM. Larger charge on the P atom is also visible.
This can give rise to the associated bright spot.}
\label{fig:Si-P_charg}
\end{figure}

Curson {\it et al.} have observed some zig-zag features in
their STM images~\cite{curson}. There is also a
bright spot associated with this feature as seen in those images.
There could be two possible origins of these, (i) P-Si hetero-dimer,
which, as already mentioned, move by $0.6$ \AA~ relative to the Si-Si
dimers, or 
(ii) the ejected Si atom in geometry {\bf I} in fig.~\ref{fig:1Pinc}.
Charge density contours in a horizontal plane
approximately $\sim 1$ \AA~ above the Si-P dimer (without the ejected Si atom
at {\bf I}) are shown in fig.~\ref{fig:Si-P_charg}. 
A zig-zag feature is distinctly visible.
There is also an excess charge density on the P atom in the displaced dimer
which can appear as a bright spot in STM. In a charge density plot
(not shown here) on a
similar plane $\sim 1$ \AA~ above the ejected Si in geometry {\bf I}, 
only the Si atom is seen and no zig-zag features, since
the surface dimer rows are not visible any more.
From these observations we conclude that the zig-zag features and
the associated bright spots can be attributed to Si-P hetero-dimers.

\subsection{$\frac{1}{4}$ ML P coverage}
\label{sec:14ML}

At $\frac{1}{4}$ ML P coverage also we study both the possibilities--P
getting adsorbed on the surface, or getting incorporated into the
surface. For each of these possibilities, we do calculations on two different
system sizes--(i) one P atom on a $(2\times 2)$ surface cell, (ii) 
four P atoms on a $(4\times 4)$ surface cell. 

\begin{figure}
\scalebox{0.30}{ \includegraphics{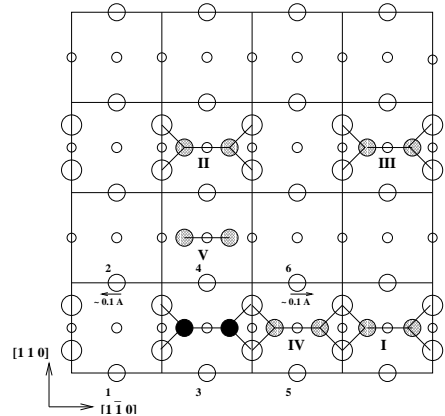} }
\caption{Initial geometries for two P-P dimers on a $(4\times
4)$ supercell studied in this work. 
Given one dimer in a position indicated by the dark
circles, five possible positions are shown for the second dimer. Movements
of two second layer Si atoms (marked {\bf 2} and {\bf 6})
when only the dark dimer is present on the surface are also shown.}
\label{fig:2dimpos}
\end{figure}

When we have four P ad-atoms adsorbed on a $(4\times
4)$ cell, they would form two P-P ortho-dimers.
Various reasonable positions of two P-P dimers 
are shown in fig~\ref{fig:2dimpos}. The BE's for the relaxed
structures starting from these
atomic configurations are given in table~\ref{table:BE2dim}. 
The dimers prefer to be separated by at least two lattice spacings (of the
square lattice on the unreconstructed Si(001) surface) along the $\langle 110
\rangle$ 
direction. In fact, there is essentially no difference in energy between 
structures {\bf II} and {\bf III}. 
This shows that there is no further energy gain in moving
the second dimer along $\langle 1\bar{1}0 \rangle$ once we have moved it along
$\langle 110 \rangle$ by two lattice spacings. On the other hand, putting two
dimers only one lattice spacing apart, either along $\langle 110 \rangle$ or
$\langle 1\bar{1}0 \rangle$ costs energy. Thus structures {\bf IV} and
{\bf V} are less favorable than
{\bf II} or {\bf III}, {\bf IV} being costlier than {\bf V}.

In order to understand why structures {\bf II} and {\bf III} 
turn out be lower in energy than 
{\bf I}, one has to look at the structure around the P-P dimer more closely.
As we have already mentioned, top layer Si-Si dimers neighboring a P-P
dimer are drawn closer to it. On the other hand, there are six second
layer Si atoms (marked {\bf 1-6} in fig~\ref{fig:2dimpos}) immediately
neighboring a P-P dimer (marked by dark circles in that figure).
Of these, two second layer atoms ({\bf 2} and {\bf 6}) 
actually move away from the P-P dimer, while the others remain
in their ideal positions as on a bare surface. When two P-P dimers are put in
structure {\bf I}, the second layer Si in between them ({\bf 6} in this case)
is frustrated since the two 
dimers tend to push it in opposite directions. This causes strain in
the structure. There is no such problem once the P dimers are separated by
at least two lattice spacings along $\langle 110 \rangle$,
in which case all second layer Si
atoms neighboring the P dimers can relax freely. 
The fact that structures {\bf II} and {\bf III} have the same energy 
also indicates that this
relaxation of the second layer Si atoms is a crucial mechanism in 
optimizing the structure. Once that has been achieved in structure
{\bf II}, 
there is no further energy gain in moving the second P dimer two lattice 
spacings along $\langle 1\bar{1}0 \rangle$.

\begin{table}[h]
\caption{BE values in eV per P atom for two P-P dimers present on a $(4\times
4)$ surface supercell of Si(001) starting from various initial atomic
geometries as discussed in the text.}
\vspace*{0.1in}
\begin{tabular}{cc}
\hline
\hline
geometry \hspace*{0.5in} & BE         \\ \hline
\hline
{\bf I} \hspace*{0.5in}      & 6.2 \\
{\bf II} \hspace*{0.5in}      & 6.3 \\
{\bf III} \hspace*{0.5in}      & 6.3 \\
{\bf IV} \hspace*{0.5in}      & 5.7 \\
{\bf V} \hspace*{0.5in}      & 5.8 \\
\hline
\label{table:BE2dim}
\end{tabular}
\end{table}

In structures {\bf IV} and {\bf V}, two P dimers are put one 
lattice spacing apart in the
initial configuration. Atomic relaxation starting from these geometries
indicate that it is energetically highly unfavorable for two P-P dimers to come
so close to each other at this coverage. 
There is a repulsion between the two dimers and they tend to
move away from each other.

Now we present our results for $\frac{1}{4}$ ML P coverage studied with one
P ad-atom on a $(2\times 2)$ surface supercell. 
We put the P ad-atom at {\bf M}, {\bf D}, {\bf H}, and {\bf C}
sites. A look at table~\ref{table:BE164ML} shows that BE's at 
{\bf M}, {\bf D} and
{\bf H} sites are essentially the same as those for an isolated
Si atom. However, binding at the {\bf C} site is significantly
stronger in the present case. At {\bf M}, {\bf D} and {\bf H}
sites, an isolated P atom is already reasonably strongly
bonded to the neighboring surface Si atoms. However, the P-Si bonding
at the {\bf C} site is rather weak, which leaves the isolated P ad-atom with 
localized electrons on it. At $\frac{1}{4}$ ML, the P atom
finds other P atoms at nearby {\bf C} sites, which gives these localized
electrons a channel to delocalize. This delocalization 
lowers the kinetic energy of
the electrons, and makes the binding stronger.
We also find that at $\frac{1}{4}$ ML coverage, two P-P
dimers on a $(4\times 4)$ surface cell have a stronger binding than a
single P ad-atom on a $(2\times 2)$ surface cell.
This shows that dimer formation by P
ad-atoms on the Si(001) surface significantly lowers their energy.

\begin{figure}[h]
\scalebox{0.45}{ \includegraphics{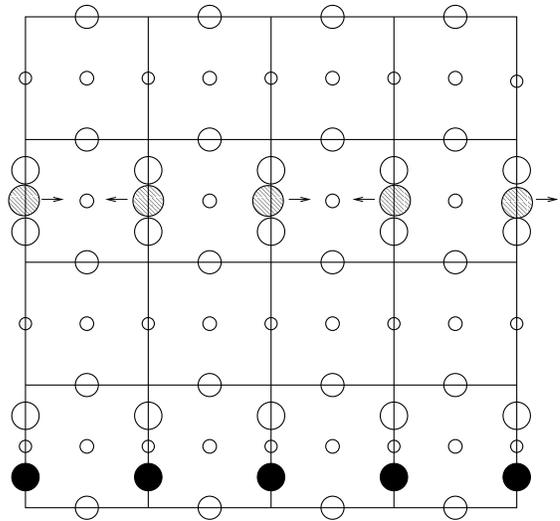} }
\caption{Starting geometry for the four ejected Si atoms (shaded circles) 
when four P atoms (dark circles) are incorporated in a $(4\times 4)$ 
surface cell.}
\label{fig:44_4Pinc}
\end{figure}

In case of P incorporation into the surface, when we have one P atom on a
$(2\times 2)$ surface cell, the ejected Si is placed in geometry {\bf I} as
explained in fig~\ref{fig:1Pinc}. The ejected Si and top four atomic layers
are relaxed. In the converged geometry, the BE is found to be 6.1 eV
per P ad-atom. While
putting four P atoms on a $(4\times 4)$ surface cell, there can be several
possibilities. However, we are guided by our calculations at $\frac{1}{8}$ ML,
where we found that two ejected Si atoms prefer to dimerize on top of
and perpendicular to surface Si dimer row. We thus incorporated all
the four P atoms in the dimers in a row of our $(4\times 4)$ cell, and put
the ejected Si atoms above the other dimer row. The starting
configuration is shown in fig~\ref{fig:44_4Pinc}. In the relaxed
geometry, the ejected Si atoms form two asymmetric dimers, as expected. The
BE turns out to be 6.2 eV per P atom. There is slight energy gain relative
to the $(2\times 2)$ cell due to dimerization of ejected Si atoms.

Now a comparison of BE's for P adsorbed and P incorporated geometries shows
that at $\frac{1}{4}$ ML, P adsorption is more favorable than P
incorporation by 0.1 eV. Thus we reach the significant conclusion 
that at a critical coverage whose value lies between
$\frac{1}{8}$ ML and $\frac{1}{4}$ ML, P atoms prefer to get
adsorbed on the surface and form P-P ortho-dimers, rather than getting
incorporated 
in the surface. Note that, while at $\frac{1}{4}$ ML P
adsorption is more favorable, in experiments, it is not surprising to find
some P incorporation concurrently.

\subsection{$\frac{1}{2}$ and full ML P coverage}
\label{sec:high}

\begin{figure}[h]
\scalebox{0.50}{ \includegraphics{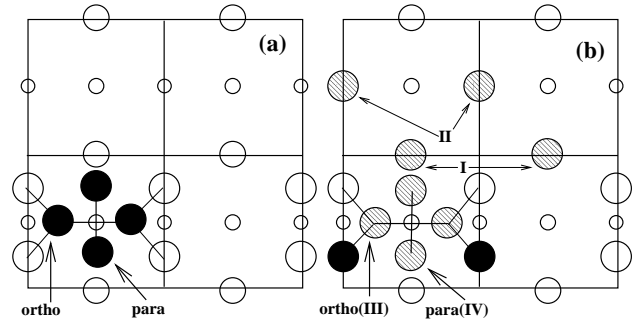} }
\caption{Two P atoms (dark circles) on a $(2\times 2)$ surface cell at 
half monolayer coverage. (a) Para- and ortho-dimer 
configurations of two adsorbed P-P dimer; (b) two P atoms
replace two surface Si atoms (shaded circles), which 
are shown in four different initial geometries.}
\label{fig:22_2Pposs}
\end{figure}

At $\frac{1}{2}$ ML also we calculated the energetics of P incorporation
into Si(001) surface, though it is expected that adsorption would be more
favorable at this coverage. In fact, that is what we find in our
calculations. In order to study adsorption at $\frac{1}{2}$ ML P coverage, 
we put two P ad-atoms in a dimerized
position on a $(2\times 2)$ surface cell. We again calculated energies
of P-P para-dimer and ortho-dimer on top of a surface dimer row. 
These two geometries are shown in
fig.~\ref{fig:22_2Pposs}(a). It turns out that the ortho-dimer is a
more favorable configuration. The P-P dimer distance is found to be
2.27 \AA~ while the Si-P distance is 2.34 \AA. 
The underlying Si dimers
are symmetrized, but the Si-Si dimer distance still remains to be 2.33 \AA.
Thus the local geometry and bonding around a P-P ad-dimer
is similar to that found around an ad-dimer in case of $\frac{1}{8}$ ML 
with a similar value for the BE. The BE turns out to be 6.3 eV per P
atom. The para-dimer configuration has a BE of 5.9 eV per P. So we
find another crucial difference between Al-Al and P-P dimers on
Si(001). Al-Al dimers showed a transition from a para- to ortho-
orientation with increasing coverage~\cite{bikash}, 
but P-P dimers always prefer an ortho- orientation.

In order to study P incorporation, we
replace two Si atoms in two dimers on a $(2\times 2)$ surface cell by
P atoms.
There are several possibilities where the ejected Si atoms 
can go. We have considered four possible arrangements for two ejected
Si atoms that are shown in Fig.~\ref{fig:22_2Pposs}(b). The
geometry {\bf I} and the ortho-dimer geometry ({\bf III}) of the Si atoms turn
out to be very close in BE.
The BE of geometry {\bf III} is 5.97 eV per P atom and that of 
geometry {\bf I} is 5.95 eV per P
atom. The para-dimer geometry ({\bf IV}) has a BE of 5.45 eV per P atom.
Geometry {\bf II} turns out to be the least favorable with a BE
of 5.12 eV per P atom.

\begin{figure}[h]
\scalebox{0.60}{ \includegraphics{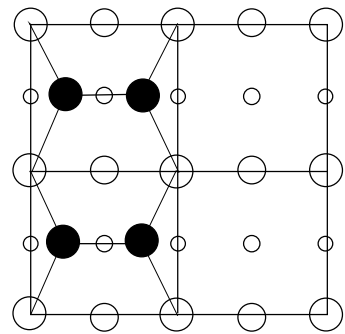} }
\caption{Four P ad-atoms (dark circles) forming two dimers on a $(2\times 2)$ 
surface cell at a full ML coverage. Reconstruction of the
underlying Si(001) surface is lifted.}
\label{fig:22_2Pdim}
\end{figure}

At one ML coverage, we put four P atoms on a $(2\times 2)$ surface
supercell. Again, a symmetric dimer-row structure of the P ad-atoms 
turns out to be the most favorable one. This is same as the
structure found for a ML As-covered Si(001). The geometry of a full ML
P covered Si(001) is shown in fig.~\ref{fig:22_2Pdim}. The P-P dimer distance
in the the relaxed structure is calculated to be 2.3 \AA. The Si-P distance
is found to be 2.38 \AA. The BE for P-P ad-dimers is found to be
6.4 eV per P atom at full ML coverage. Another important feature seen
at this coverage is that the reconstruction of the underlying Si
surface is lifted, just as was found for As on Si(001).
We have seen %at $\frac{1}{4}$ ML 
that it is unfavorable for two P-P dimers to come too
close to each other on the surface. Some of this 
strain could be released by missing P-P dimer rows as found in experiments.
However, since we are studying full ML coverage with a $(2\times
2)$ surface supercell, we cannot explore this possibility in our
calculations.

Our findings that at higher coverages P adsorption becomes more
favorable, and that P ad-atoms prefer to form 
dimers, are consistent with the conclusions reached by 
Yu {\it et al.} and Wang {\it et al.}~\cite{yu3,wang}.

\section{Conclusions}

We have done a systematic first-principles pseudopotential density
functional study of structural changes induced by P on Si(001). For
adsorption of an
isolated P atom, the {\bf M} site turns our to be energetically most
favorable. However, up to a P coverage of $\frac{1}{8}$ ML, replacement of
surface Si atom by P is even more beneficial energetically due to the
participation of the {\bf M} site. The
resulting Si-P hetero dimers give rise to the zig-zag and associated
bright features in STM images. The ejected Si atoms prefer to form
dimer chains perpendicular to the surface dimer rows. At some critical
coverage between $\frac{1}{8}$ ML
and $\frac{1}{4}$ ML, adsorption of P becomes more favorable than
incorporation into the surface. At all coverages, P-P ortho-dimers on
top of Si dimer rows are more favorable than para-dimers. This is in contrast
to Al-Al dimers on Si(001) which prefer to reside 
between surface dimer rows, and
show a transition from an para- to ortho- orientation with increasing
coverage. At $\frac{1}{2}$ ML coverage P ad-atoms also form dimers. At a full
ML coverage P atoms show a propensity to form dimer rows after
lifting reconstruction of the Si surface. There could
be some missing dimer rows to relieve strain in the system.

\label{sec:conclude}

\end{document}